\documentclass{article}
\usepackage{meta}
\usepackage{amssymb}
\usepackage{amsmath}
\usepackage{epsfig}
\title{$\mathcal{PT}$-Symmetric Dimers with Time-Periodic Gain/Loss Function} 
\makeatletter
\def\name#1{\gdef\@name{#1\\}}
\makeatother
\name{{\bf \large  Demetra Psiachos$^{1}$, Nikos Lazarides$^{1,2,*}$, G.P. Tsironis$^{1,2}$}}
\address{
$^1$Crete Center for Quantum Complexity $\&$ Nanotechnology, Department of Physics, 
University of Crete, \\
P. O. Box 2208, 71003 Heraklion Greece \\
$^2$Institute of Electronic Structure and Laser, Foundation for Research and Technology-Hellas, \\
P. O. Box 1527, 71110 Heraklion Greece \\
*corresponding author, E-mail: {\tt nl@physics.uoc.gr}
}
\begin{document}
\maketitle
\begin{abstract}
$\mathcal{PT}$-symmetric dimers with a time-periodic gain/loss function in a balanced
configuration where the amount of gain equals that of loss are investigated analytically
and numerically. Two prototypical dimers in the linear regime are investigated: a system 
of coupled classical oscillators, and a Schr\"{o}dinger dimer representing the coupling 
of field amplitudes; each system representing a wide class of physical models.
Through a thorough analysis
of their stability behaviour, we find that turning on the coupling parameter in the
classical dimer system, leads initially to decreased stability but then to re-entrant
transitions from the exact to the broken $\mathcal{PT}-$phase and vice versa, as it is 
increased beyond a critical value. On the other hand, the Schr\"{o}dinger dimer behaves
more like a single oscillator with time-periodic gain/loss. In addition, we are able 
to identify the conditions under which the behaviour of the two dimer systems coincides
and/or reduces to that of a single oscillator.
\end{abstract}

\section{Introduction}
Artificial materials with engineered properties have recently attracted a lot of
attention. A well-known paradigm is the parity-time (${\cal PT}$) symmetric systems,
whose properties rely on a delicate balance between gain and loss. The ideas and notions 
of ${\cal PT}-$symmetric materials, which do not separately obey the parity ($\cal P$) 
and and time ($\cal T$) symmetries, instead exhibiting a combined ${\cal PT}$ symmetry,
have their roots in non-Hermitian quantum mechanics \cite{Hook2012}. 
The theory has been extended to optical lattices \cite{ElGanainy2007,Makris2008},
and the predicted ${\cal PT}-$symmetry breaking as well as other properties have been 
experimentally observed in optical 
\cite{Guo2009,Ruter2010,Regensburger2012,Feng2012,Feng2014} and microresonator 
\cite{Peng2014} systems.
The theory subsequently evolved to include nonlinear lattices \cite{Dmitriev2010}
and oligomers \cite{Li2011,Ambroise2012}. The application of these ideas in electronic 
circuits \cite{Schindler2011,Bender2013}, provides a convenient  platform for testing 
${\cal PT}-$related ideas within the framework of easily accessible experimental
configurations. Also, the construction of ${\cal PT}-$metamaterials, reliant
on gain and loss-providing components, has been recently suggested 
\cite{Lazarides2013a,Tsironis2014}. 
${\cal PT}-$symmetric systems can migrate from the exact into the broken phase as the 
parameters of the system are varied.
In the exact phase, ${\cal PT}-$symmetric systems have real eigenvalues, and thus 
propagating modes as well as bound states, despite being non-Hermitian.
One essential difference from 
conventional systems is that the total energy, instead of being conserved, oscillates
because the eigenmodes are not orthogonal. Another consequence of the non-orthogonality
of the eigenmodes is non-reciprocal propagation \cite{Ruter2010}, an asymmetry in the
propagation in the two channels, even as the initial conditions are reversed.
In the broken phase, the energy increases exponentially, and the propagation is 
unstable: amplified under gain or decaying with loss. The band structure of the 
system in the broken phase acquires imaginary components for some values of the wavevector
\cite{Makris2008,Ruter2010,Regensburger2012}.
Experimentally, grating structures have been synthesized and used to demonstrate 
greatly enhanced or reduced reflection in the broken phase \cite{Regensburger2012,Feng2012},
paving the way for more extensive studies of these remarkable phenomena, and later,
for the design of new devices exploiting them. Recently, a ${\cal PT}-$symmetric
coherent perfect absorber was realized which can completely absorb light in its
exact phase \cite{Sun2014}.

In the present work, two different ${\cal PT}-$symmetric models are investigated:
a system consisting of two coupled linear oscillators (classical dimer),
and a linear Schr{\"o}dinger dimer (LSD). Both systems include terms providing 
gain and loss, which render them ${\cal PT}-$symmetric.   
In most previous works' implementation, the gain/loss parameter remains constant
in time, although recently a time-varying gain/loss function has been implemented
in coupled fibre loops \cite{Regensburger2012}. 
More generalized ${\cal PT}-$symmetric dimer models in various contexts, that include 
cubic or other nonlinearities and constant gain/loss functions, have been also
investigated theoretically 
\cite{Ramezani2010,Benisty2011,Barashenkov2012,Lupu2013,Duanmu2013,Cuevas2013,Barashenkov2013,Molina2014,Alexeeva2014}.
Moreover, the addition of a second spatial dimension in the nonlinear Schr{\"o}dinger 
dimer, which activates spatial dispersion, results in the appearence of solitons and
rogue waves as in ${\cal PT}-$symmetric dual-core waveguides \cite{Bludov2012,Bludov2013}.
In extended nonlinear ${\cal PT}-$symmetric lattices, on the other hand, the existence 
of localized excitations in the form of discrete solitons \cite{Konotop2012} and  
discrete breathers \cite{Lazarides2013a} has been demonstrated.  
In the present work, a periodic in time  
gain/loss function having an opposite effect on the two components is considered.
When averaged over one period, the amount of gain equals the loss in each component.
This time-dependence of the gain/loss function introduces yet another parameter into
the problem, and this results in multiple ${\cal PT}$ transitions from the exact phase,
where stable propagation exists, to the broken phase, where all solutions diverge.
In the following we describe the two above-mentioned systems and discuss the 
methods used in their analysis. The next section is devoted to the ${\cal PT}-$symmetric
classical dimer, while a description of the ${\cal PT}-$symmetric LSD is given in Section 3. 
The conclusions are presented in the last section.

\section{Linear $\mathcal{PT}$-Symmetric Classical Dimer}
\label{sec:classical}
A $\mathcal{PT}$-symmetric dimer comprising two linear classical oscillators with a 
time-periodic gain/loss function  $\gamma=\gamma(t)$ with $t$ being the temporal variable,
is modeled by the equations
\begin{align} 
\label{kdimer1}
\ddot{x}_1+2\gamma(t)\dot{x}_1+\omega_0^2x_1&=k(x_2-x_1) \\ 
\label{kdimer2}
\ddot{x}_2-2\gamma(t)\dot{x}_2+\omega_0^2x_2&=k(x_1-x_2),
\end{align}
where $x_1$ and $x_2$ are the longitudinal-mode displacements from equilibrium of the 
first and second oscillator, respectively, $\omega_0$ is their (common) eigenfrequency,
$k$ is the coupling constant, and the overdots denote differentiation with respect to
time $t$. Note that the commonly-appearing damping factor, which acts on the velocity 
of each oscillator, has been replaced by the gain/loss function $\gamma(t)$;
also note that the signs in front of $\gamma(t)$ are opposite in the two equations.

A particularly simple form of $\gamma(t)$ which allows the reduction of Eqs. (\ref{kdimer1})
and (\ref{kdimer2}) to an area-preserving mapping for discrete time is the following 
piece-wise linear gain/loss function
\begin{equation}        
 \gamma (t) = \left\{ \begin{array}{ll}
        +\gamma_0 & \mbox{if $0 \le t<\tau_1$};\\
        -\gamma_0 & \mbox{if $\tau_1 \le t< \tau_2$}\end{array} \right.            
\label{gainloss}
\end{equation}
where $\tau_1 +\tau_2\equiv T$, the period of $\gamma(t)$, and the constant $\gamma_0$
is defined as positive real number ($\gamma_0 >0$). Using Eq. (\ref{gainloss}),
we have, say for the first oscillator, loss in the first part of the cycle ($\tau_1$ 
time units) and gain in the second part of the cycle ($\tau_2$ time units).
With this form of $\gamma (t)$, Eqs. (\ref{kdimer1}) and (\ref{kdimer2}) are easily 
solvable in each part of the cycle separately. 
Note that for the dimers having an averaged balanced configuration, $\tau_1$ and $\tau_2$ 
need not be equal; in the present work however, the case of $\tau_1 =\tau_2 =\tau$ is 
investigated.
The solution of the $x_i$, $i=1,2$, in each half-period $\tau\equiv T/2$ of $\gamma(t)$ 
is of the form:
\begin{equation}
x_i(t)=c_1\mathrm{e}^{r_1t}+c_2\mathrm{e}^{r_2t}+c_3\mathrm{e}^{r_3t}+c_4\mathrm{e}^{r_4t}
\label{xi}
\end{equation}
for generally complex constants $r_j$, $c_j$, $j=1,...,4$, that can be expressed
analytically as a function of the initial conditions (at time $t=0$).
Then, the coefficients of $x_1 (t=0) \equiv x_{1,0}$, $\dot{x}_1 (t=0) \equiv \dot{x}_{1,0}$,
$x_2 (t=0) \equiv x_{2,0}$, and $\dot{x}_2 (t=0) \equiv \dot{x}_{2,0}$
are collected and used to construct the propagation matrix
$M_{0\rightarrow \tau}$ ($M_{\tau \rightarrow T}$) for the first (second) half
of the period $T$ of the gain/loss function, using the same method as in Ref. \cite{Tsironis2014}. 
The relation of the amplitudes and
the velocities at time $t=0$ with those at time $t=T$ can then be written in the
matrix form
\begin{equation}
\begin{pmatrix}
  x_1       \\
  \dot{x}_1 \\
  x_2       \\
  \dot{x}_2
\end{pmatrix}
=\underline{\underline{M}}_{0\rightarrow \tau}(\tau)
 \underline{\underline{M}}_{\tau \rightarrow T}(\tau)
=\underline{\underline{M}}_{G/L}(\tau)
\begin{pmatrix}
  x_{1,0}       \\
  \dot{x_{1,0}} \\
  x_{2,0}       \\
  \dot{x_{2,0}}
\end{pmatrix}
\label{Mmatrix}
\end{equation}

\begin{figure}[t!] 
\includegraphics[width=8cm]{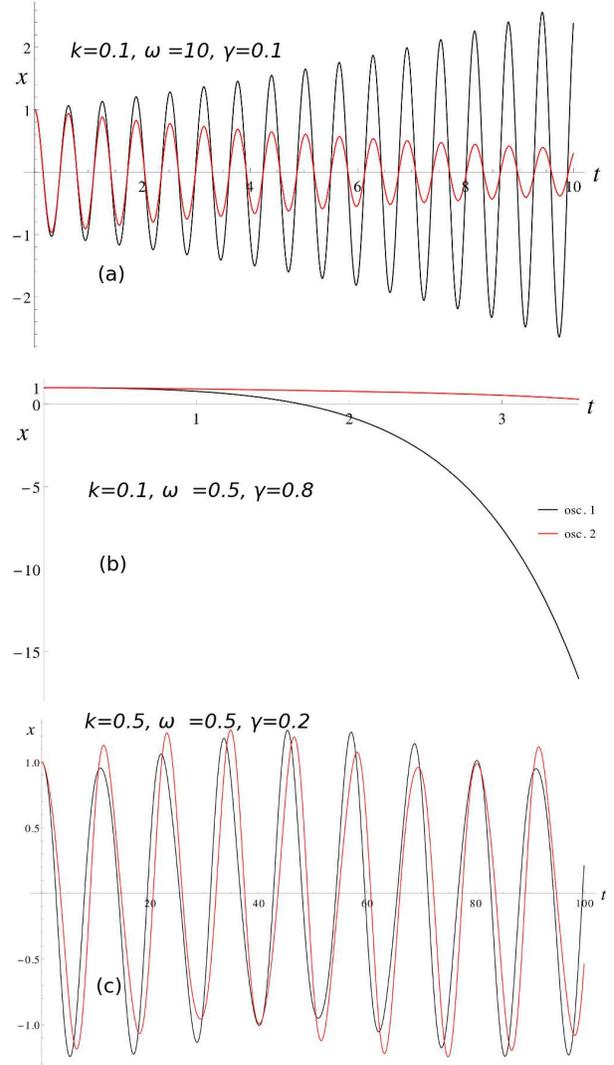}
\caption{(Color online)
Oscillation amplitude for constant in time gain/loss function $\gamma$,
for the initial conditions $(x_1,\dot{x}_1,x_2,\dot{x}_2)=(1,0,1,0)$, and
(a) $\omega_0>\gamma$ (both oscillators underdamped);
(b) $\omega_0<\gamma$ (both oscillators overdamped);
(c) $\omega_0<\gamma$.
In this case the first oscillator is subject to gain and the second is
subject to loss for all times. The motion is unstable in (a) and (b),
while it is stable in (c).
}
\label{ampl}
\end{figure}

\subsection{Constant gain/loss function}
\label{sec:constgl}
\begin{figure}[t!] 
\includegraphics[width=8cm]{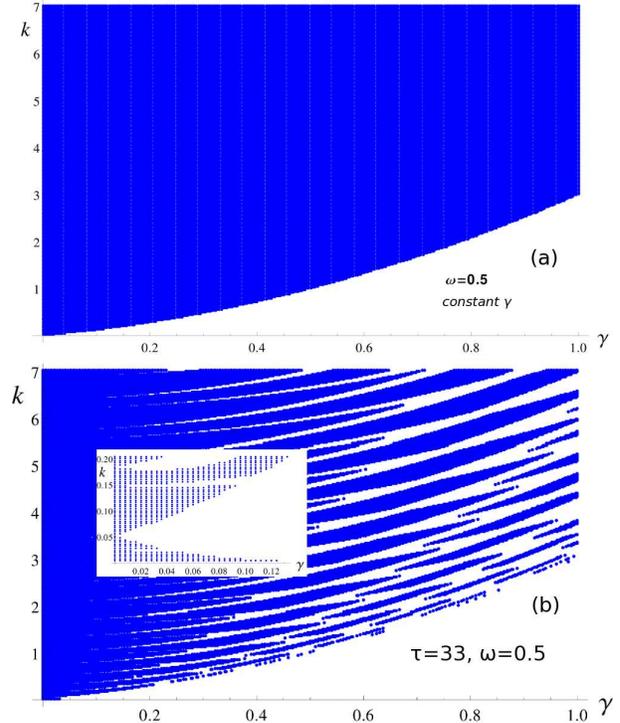}
\caption{Stability phase diagram on the $k-\gamma$ parameter space for 
(a) constant gain/loss function $\gamma$; 
(b) time-dependent $\gamma= \gamma(t)$ with half-period $\tau=33$.
The inset in (b) shows the stability in the low-$k$ region which is not
present in the case of constant $\gamma$.
The shaded (blank) regions correspond to stable (unstable) oscillations.
}
\label{kgammaG}
\end{figure}

For constant $\gamma$ the amplitudes of the oscillators for various cases
are shown in Fig.~\ref{ampl}. Both oscillators start with amplitude equal to
unity and zero velocity, and the first oscillator is considered to have gain 
while the second one has loss. The variable $\tau$ here is infinity and the 
horizontal shows the real time. Although such a characterization is strictly
valid only for $k=0$, the first case, (a), where $\omega_0>\gamma$, is termed
`underdamped' in order to connect the behaviour to the single-oscillator problem
studied in~\cite{Tsironis2014}, where the effective frequency, modified by $\gamma$,
is real and leads to oscillatory motion, while case (b), where $\omega_0<\gamma$,
is called `overdamped', shows monotonic behaviour. 
In case (c) the motion is marginally-stable as it is stabilized by a sufficiently
large $k$. Regions of stability as a result of nonzero $k$ can be found in either
$\omega_0>\gamma$ or $\omega_0<\gamma$. This is not only clear from the solutions
themselves, but from an analysis of the eigenvalues of the propagation matrix 
$M_{G/L}$, as will be discussed later in Sec.~\ref{sec:kstab}.
The stability behaviour under constant gain/loss has already been discussed in
Ref.~\cite{Schindler2011}, where a system very similar to that described by Eqs. 
(\ref{kdimer1}) and (\ref{kdimer2}) was studied theoretically, and it was implemented
using electric circuits containing active-gain/loss components. In that study,
the ratio of mutual to self-inductances of the two components comprising the
$\mathcal{PT}$-symmetric dimer is related to the coupling $k$ here. However, 
no quantitative comparisons can be made between the two models due to different
parametrization. In our model, Eqs. (\ref{kdimer1}) and (\ref{kdimer2}), from the 
analysis of the eigenfrequencies the stability condition for $\gamma$ constant in 
time is found to be
\begin{equation}
\label{condition1}
  \gamma \leq 
  \sqrt{\frac{1}{2}\left(k+\omega_0^2\right)-\frac{1}{2}\sqrt{2k\omega_0^2+\omega_0^4}} .
\end{equation}
A phase diagram showing how nonzero $k$ stabilizes the motion in the constant $\gamma$ 
case is shown in Fig.~\ref{kgammaG}a. The limit $k\rightarrow 0$ in the constant $\gamma$
system gives the result of instability except for $\gamma=0$, although this is a fictitious
result due to the ambiguities inherent in the mapping of the dimer onto an uncoupled
system (Sec.~\ref{sec:intra}).

\subsection{Time-periodic gain/loss and stability criterion}
\label{sec:kstab}
When the gain/loss function varies in time, there are changes in the stability 
behaviour. We have a reduced phase space for stability in the `underdamped' region
$\omega_0>\gamma$ and an increased phase space in the `overdamped' region 
$\omega_0<\gamma$. Fig.~\ref{kgammaG}, shows an example of the stabilizing effect 
of the coupling parameter $k$, something also noted experimentally in Ref.
\cite{Regensburger2012}. That study examined the $\mathcal{PT}$ threshold of 
two coupled waveguides subject to an opposing alternating $\gamma(t)$
-albeit with a fixed half-period $\tau$- while the real part of the $\mathcal{PT}$ potential, 
implemented as a time-varying phase factor, was adjusted. Unlike in the 
present model, their coupling parameter was also made to vary in time. 
Fig.~\ref{kgammaG}b shows how introducing a time variation in $\gamma$ results
in decreased stability in some regions and increased stability in others
as compared with the case of constant $\gamma$ (inset).
\begin{figure}[htb] 
\includegraphics[width=8cm]{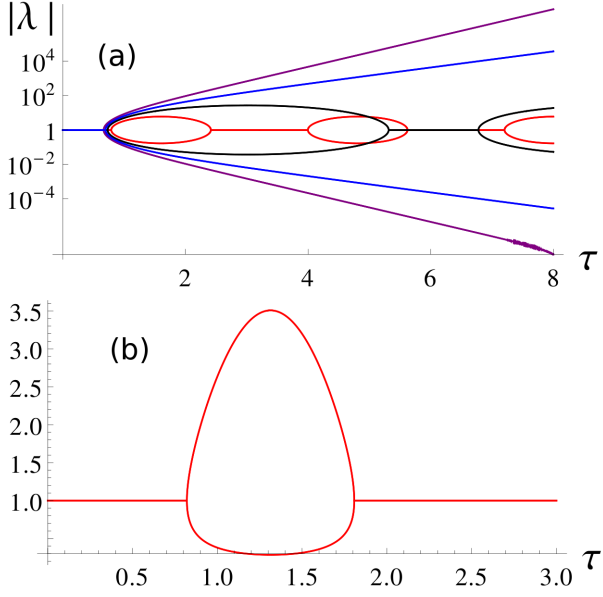}
\caption{Magnitudes $\left|\lambda\right|$ of the four eigenvalues 
as a function of the gain-loss oscillation half-period $\tau$
(a) for $k=0$ and $\omega=1.4$. 
Purple and Blue curves respectively: $\gamma=1.7$ and 1.5 (overdamped), 
black and red curves: $\gamma=1.3$ and 1 respectively (underdamped). 
Two eigenvalues drop below and two go above unity in the unstable regions.
There is a qualitative correspondence to the behaviour of the eigenvalues
in the overdamped and underdamped cases of linear Schr\"{o}dinger dimer 
system, where $V$ replaces $\omega_0$.
In (b), the parameters are $k=0.11$, $\omega_0$=1.4, and $\gamma=0.8$.
} 
\label{eigs}
\end{figure}

The stability points (marginally-stable oscillations for $x_1(t)$ and $x_2(t)$) 
occur when the (complex) eigenvalues of $M_{G/L}$ all simultaneously have 
the absolute value of unity. There was no other type of stability present;
for no parameters were the eigenvalues simultaneously less than unity,
which would have implied asymptotic stability (decaying to zero).
In the regions of instability, two eigenvalues drop below while the other two
go above unity (Fig.~\ref{eigs}). Such oval-like structures appear in other 
studies of stability in $\mathcal{PT}$ systems~\cite{Makris2008,Schindler2011}.
The instability onset is not as sharp as Fig.~\ref{eigs} seems to indicate;
the motion will be bounded by some larger magnitudes than the initial value
for values of $\tau$ below or above the unstable region, but the oscillation 
will still be bounded.
\begin{figure}[t!] 
\includegraphics[width=7cm]{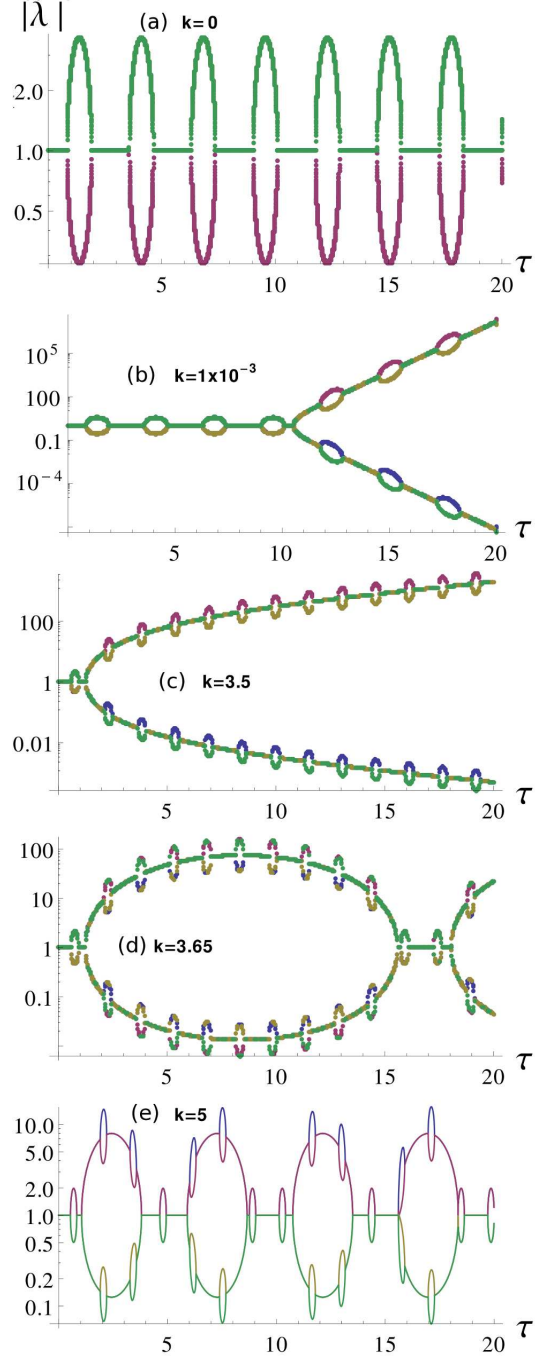}
\caption{Magnitudes of the four eigenvalues for the propagation matrix $M_{G/L}$
in Eq.~\ref{Mmatrix} for different values of the coupling $k$, as $\tau$  is
varied. The values of $\omega_0$ and $\gamma$ are fixed at 1.4 and 0.8 respectively.} 
\label{kvar}
\end{figure}

\subsection{Effect of intra-dimer coupling}
\label{sec:intra}
The full analytical expansion for the eigenvalues of $M_{G/L}$ in Eq.~\ref{Mmatrix} 
is extremely complex while series expansions yield little insight as due to the 
correlation of the different variables, the convergence properties of any 
expansion are very complicated.
If the limits leading to Fig.~\ref{kgammaG}, where $k=0$ appears to lead to 
instability, are taken in reverse, with the limit $\tau\rightarrow\infty$
of the $k=0$ system taken, the result is indeterminate:
neither stable nor unstable as discussed later at
the beginning of Sec.~\ref{sec:intra} and as seen in Fig.~\ref{kvar}a and 
Fig.~\ref{klinlog}. Lastly, a prediction of total stability for $\omega_0>\gamma$
and instability for $\omega_0<\gamma$ arises if a path such as that in 
Fig.~\ref{schrophaseconst}c$\rightarrow$a,
where $\gamma$ is fixed while $\omega_0$ is varied, is taken. Essentially, 
by choosing such a path, the periodicity with variable $\omega_0$ of the stripes
reduces to zero, yielding the identity operation.

Likewise, the single oscillator with constant nonzero $\gamma$
is everywhere unstable, but performs according to Fig.~\ref{schrophaseconst}a
if $\tau$ is made to approach infinity while $\omega_0$ is varied for fixed $\gamma$, again
because the periodicity of the $\mathcal{PT}$ symmetry-breaking stripes reduces to zero. On 
the other hand, the approach $\tau\rightarrow\infty$ for fixed $\omega_0$ and $\gamma$ 
is indeterminate as in Fig.~\ref{kvar}a. 
Thus, we may reconcile the $k=0$ limit of the dimer
with the behaviour of the single oscillator if the limits are chosen in a particular sequence.

The approaches $k\rightarrow 0,$
$\tau\rightarrow \infty,$ and $\gamma\rightarrow\omega_0$ are correlated amongst 
themselves and a mapping for a consistent (path-independent) unified approach
from the dimer to the single-oscillator is currently being sought. Even numerically,
the reduction of the coupled system to one of the phase diagrams of the single-oscillator is not
valid for all paths towards the double limit $k\rightarrow 0$,
and $\tau\rightarrow \infty,$ which is essentially undefined. Keeping this
in mind, we explain below exactly how our calculations were performed for the limiting cases.

\subsubsection{Zero coupling limit, $k$=0}
In studying the case of $k=0$ or no coupling between oscillators, we choose to
refer to the solution of a single oscillator~\cite{Tsironis2014}, because the
system of two oscillators does not in general reduce to that of one oscillator
in the limit $k\rightarrow 0$, as it depends on the path taken. In the case of a 
single oscillator, there is a periodic set of stable trajectories for the underdamped
limit $\omega_0>\gamma$ for all values of $\tau$. 
In this limit, the onset and/or end of the instabilities occurs periodically where
\cite{Tsironis2014}
\begin{equation}
  \cos{(2\delta \tau_{crit})}=\left( \frac{2\gamma^2}{\omega_0^2}-1\right),
  \label{instability}
\end{equation}
where $\delta=\sqrt{\omega_0^2-\gamma^2}$.
For the overdamped ($\omega_0<\gamma$) case, 
there is only one solution to Eq.~\ref{instability}: 
$\tau_{crit}=\arccos{(\gamma/\omega_0)}/\delta$, and the bifurcation in the magnitude 
of the eigenvalues is permanent beyond this point as it does not close in over itself
(purple and blue curves in Fig.~\ref{eigs}a). For values of $\gamma$ below $\omega_0$
the bifurcations close in over themselves and periodically repeat following 
Eq.~\ref{instability}, while the width and height of the unstable regions decreases
with the value of $\gamma/\omega_0$. Finally, in the limit $\omega_0\gg\gamma$
the splitting of the eigenvalue magnitudes in the unstable regions (size of the bubbles) 
is reduced to zero whereupon we recover the undamped limit of conventional stability.
\begin{figure}[h] 
\includegraphics[width=8cm]{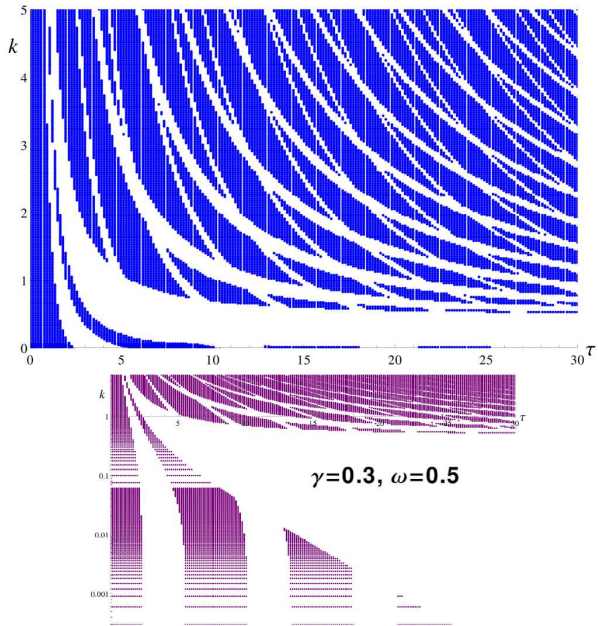}
\caption{Stability phase diagram on the $k-\tau$ parameter plane for the 
linear classical dimer for constant $\gamma=0.3$ and $\omega_0=0.5.$ The
lower plot is the same as the upper plot but with the vertical axis using 
a log scale in order to highlight the low$-k$ regime.
The shaded (blank) regions correspond to stable (unstable) oscillations.}
\label{klinlog}
\end{figure}

Note that Eq.~\ref{instability},
for constant $\gamma$, or infinite $\tau$, can have a $\mathcal{PT}$-symmetry breaking
transition across the line ${\omega_0}_{crit}=\gamma$ so the system can indeed
reduce to the behaviour in Fig.~\ref{schrophaseconst}a if the paths
are chosen according to prescription outlined at the start of this section, of
varying $\omega_0$ while holding the other parameters fixed, so that in Eq.~\ref{instability}
we have $\tau$ instead of $\tau_{crit}$ and ${\omega_0}_{crit}$ instead of ${\omega_0}$. On 
the other hand, the stability prediction of Eq.~\ref{instability} 
is indeterminate for $\tau\rightarrow\infty$ (see \textit{e.g.} Fig.~\ref{kvar}a and \ref{klinlog}
for $k=0$) as $\tau$ is varied.
In addition to the ${\omega_0}_{crit}\rightarrow\gamma$ case of Eq.~\ref{instability},
a separate condition can be derived for when the limit $\omega_0\rightarrow\gamma$ 
(system parameters as opposed to phase boundary as in the previous paragraph)
is taken at the outset. If this is done, then we have a $\mathcal{PT}$ symmetry-breaking point
occurring at $\tau_{crit}=1/\gamma$. This is the cause of the peak which appears 
for small $\tau$ in Figs.~\ref{schrophaseconst} and \ref{schrophase}.

\subsubsection{Nonzero $k$}
The effect of nonzero $k$ is to shorten the region in $\tau$ where 
periodic stability/instability regions can occur. This is clearly seen by comparing
Figs.~\ref{kvar}a-c. However, as $k$ becomes large, 
beyond the critical value which occurs between Fig.~\ref{kvar}c and d,
a new regime is reached: that of a very noisy but infinitely-periodic region of 
$\mathcal{PT}$-symmetry
breaking and recovery (Fig.~\ref{kvar}d-e). Similar to the $k=0$ case
of ${\omega_0}_{crit}=\gamma+0^+$ where the bifurcation closes in on itself
to produce an infinite series of $\mathcal{PT}$-symmetry breaking and recovery, here also, for
a critical value $k=k_{crit}$, between the cases pictured in Fig.~\ref{kvar}c and d, 
the bifurcation behaves in a similar manner, leading to the production of a 
more-complex series of 
superimposed bubbles. The small, more numerous bubbles, are due to the underdamping 
$\omega_0>\gamma$ while the large ones are due to $k>k_{crit}$.

Just as the $\omega_0\gg \gamma$ case for the uncoupled oscillators, depicted in the
phase plots of Fig.~\ref{schrophaseconst}, leads to increased stability, or a longer
$\tau_{crit}$ (Eq.~\ref{instability}), here also, if we have infinitely-large $k$, 
then we will recover the regime of continuous, unbroken symmetry because
the symmetry-breaking bubbles shrink to zero size. 
In addition, we see from Fig.~\ref{schrophaseconst}
that increasing $\tau$ leads to greater stability for an underdamped system, as we can 
otherwise see from Fig.~\ref{klinlog}
where the density of stable regions increases for larger $\tau$. In 
the regime $\tau\rightarrow\infty$ and large $k$ the classical dimer system behaves as 
independent \textit{undamped} harmonic oscillators, showing conventional rather than
$\mathcal{PT}$-symmetric stability. Studying the overdamped, $\omega_0<\gamma$, system
is of little interest as the $\mathcal{PT}$-symmetric region is confined to a smaller
phase space in comparison with the underdamped system as the recoveries for $k$ larger 
than the critical value have only the larger $k$ bubbles since the smaller ones do not exist.
Thus, while the plots in Fig.~\ref{kvar} reveal the mechanism (through the bifurcations 
closing in on themselves) by which $k$ influences the size of
the region of periodic $\mathcal{PT}$-symmetry breaking and recoveries, the $k-\tau$ phase 
diagram in 
Fig.~\ref{klinlog} depicts the overall result over all $k$ values. In both of these examples, 
Fig.~\ref{kvar}
and ~\ref{klinlog}, with $\gamma<\omega_0,$ the qualitative behaviour is the same, 
even as the exact
parameters differ. We find an infinite series of $\mathcal{PT}$-symmetry breaking/recoveries 
for $k\rightarrow 0$ while, as seen in 
Figs.~\ref{kvar}a-c, the $\mathcal{PT}$-symmetry breaking/recovery
range in $\tau$ is reduced as $k$ is increased, something which is especially evident from 
the lower plot of Fig.~\ref{klinlog}, until there is just one recovery (\textit{viz.} 
large gap in the phase
plot). Upon increasing $k$ further, the infinite series of $\mathcal{PT}$-symmetry breaking/recoveries
is regained beyond a critical value of $k$.

\section{Linear Schr\"{o}dinger $\mathcal{PT}$-Symmetric Dimer}
Systems of Schr\"{o}dinger dimers have been studied widely in the context of optics, 
where $\mathcal{PT}$-symmetry phenomena can be realized experimentally
using e.g. gratings or coupled waveguides providing alternating gain/loss.
The dynamical behaviour of the Schr\"{o}dinger $\mathcal{PT}$-symmetric
dimer is governed by the equations
\begin{eqnarray} 
  \label{wfdimer1}
    i\dot{\Psi}_a &=-i\gamma\Psi_a +V\Psi_b   \\ 
  \label{wfdimer2}
    i\dot{\Psi}_b &=+i\gamma\Psi_b +V\Psi_a  
\end{eqnarray}
where $\Psi_i$ is the wavefunction at site $i$ ($i=a,b$), with $V$ being the
coupling between the two sites,
$\gamma=\gamma(t)$ is the gain/loss function given by Eq. (\ref{gainloss}),
and the overdot denotes derivation with respect to the temporal variable.
The (non-Hermitian) Hamiltonian used to derive Eqs. (\ref{wfdimer1}) and
(\ref{wfdimer2}) is given by
\begin{equation}
\mathcal{H}=\begin{pmatrix}
-i\gamma&V\\
V&i\gamma
\end{pmatrix}.
\label{hschro}
\end{equation}
For $\gamma=0$, Eqs. (\ref{wfdimer1}) and (\ref{wfdimer2}) form a completely
integrable system; thus, the Hamiltonian $\mathcal{H}$ and the total
probability (norm) 
\begin{equation}
\label{totalprob}
   P=\sum_{i=a,b} \left|\Psi_i \right|^2 ,
\end{equation}
are conserved.
For $\gamma \neq 0$ neither $P$ nor $\mathcal{H}$ are conserved but instead,
they oscillate in time; however, a more complex conserved quantity still
exists \cite{Pickton2013}.

\subsection{Constant gain/loss}
In contrast with the case of classical oscillators (\textit{e.g.} the single one
or the dimer as in Fig.~\ref{ampl}), where a constant $\gamma$ in the broken 
$\mathcal{PT}$-symmetry phase led to either underdamped or overdamped growth or decay,
the growth or decay here is always monotonic. Indeed, in experiments on light
intensity in coupled waveguides~\cite{Ruter2010}, it has been found that above
the $\mathcal{PT}$ transition point the light propagates, exponentially amplified,
through one particular waveguide irrespective of the initial conditions.
\begin{figure}[h]
\includegraphics[width=8cm]{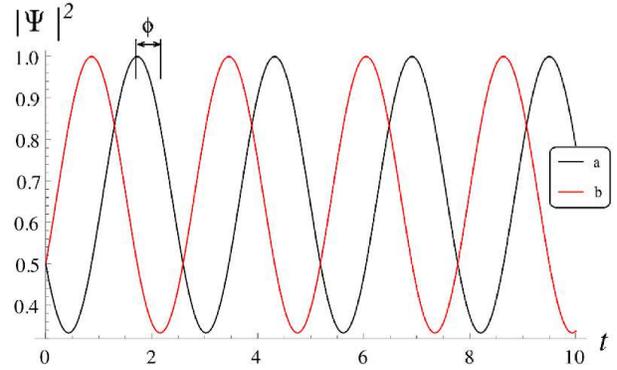}
\caption{Magnitudes of the linear Schr{\"o}dinger dimer wavefunctions for a 
constant nonzero damping $\gamma =0.7$, where
oscillator $a$ is subject to gain and oscillator $b$ is subject to loss, 
for the initial
conditions $(\Psi_a,\Psi_b)=(1/\sqrt{2},1/\sqrt{2})$ and $V$=1.4. The phase 
shift $\phi$,
with respect to the phase difference $\pi$ corresponding to $\gamma=0,$ is shown.}
\label{phaseGL}
\end{figure}

\begin{figure}[h]
\includegraphics[width=8cm]{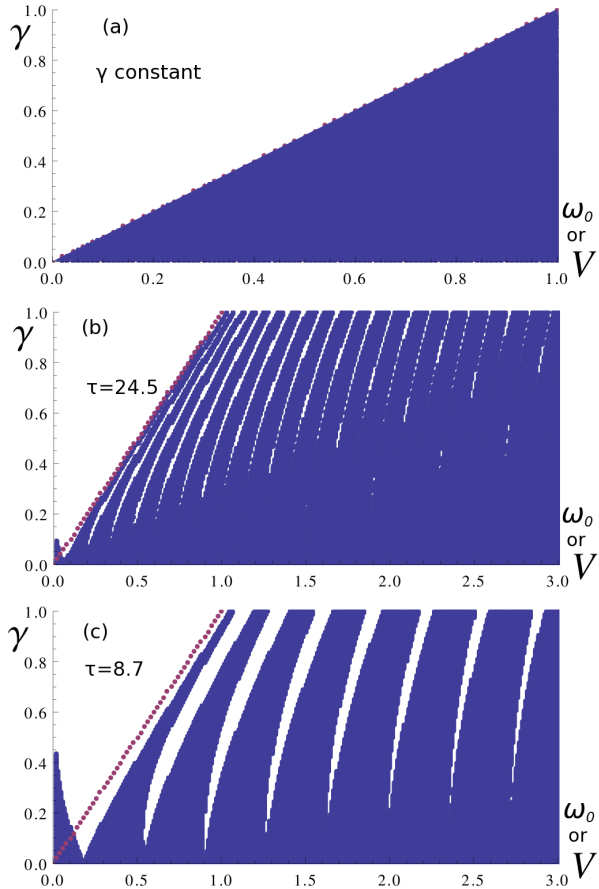}
\caption{Stability phase diagram on the $\gamma-(V$ or $\omega_0)$ parameter 
plane for the linear Schr\"{o}dinger dimer, for 
(a) constant $\gamma$ and (b)-(c) oscillating $\gamma(t)$.
Plots (b) and (c) may also correspond to the linear classical dimer with $k=0$
or to the single classical oscillator.
(a) is achieved for the classical systems only in a specific regime as
discussed in the text; otherwise it is blank. 
The shaded (blank) regions are stable (unstable). The purple dotted line shows the 
line $\gamma=V$ or $\omega_0$ which corresponds to the stable/unstable boundary in (a).}
\label{schrophaseconst}
\end{figure}

For $V>\gamma$ the system is always $\mathcal{PT}$-symmetric (Fig.~\ref{schrophaseconst}a), 
in stark contrast with the classical oscillator or dimer with $k=0$ where the theory
predicts complete instability when the limit $\tau\rightarrow\infty$ is taken first
(start of Sec.~\ref{sec:intra}), and the magnitudes of the wavefunctions 
$\Psi_a(t)$ and $\Psi_b(t)$ of the coupled Schr\"{o}dinger dimer undergo stable 
oscillations as they do in the exact $\mathcal{PT}$-symmetric phase of the classical
dimer under constant gain/loss considered in Sec.~\ref{sec:constgl}.
For $\gamma=0$ the wavefunction magnitudes are phase shifted with respect to each other 
by $\pi$ while when $\gamma\rightarrow V+0^-$ the phase difference asymptotically
approaches zero. A phase shift of $\pi/2$ occurs at $\gamma=V/\sqrt{2}$.
Fig.~\ref{phaseGL} shows an example of the phase difference between the two dimer
components. Such phase delays between two eigenmodes of a coupled two-channel waveguide
have been observed in experiments as $\gamma$ is varied~\cite{Ruter2010} and they are
attributed to the nonorthogonality of the modes. This constant phase delay is not
present in the classical dimer system because non-zero $k$ is necessary for stability
when the damping factor is constant; for nonzero $k$, as seen in Fig.~\ref{ampl}c, the
signals are chirped and eventually coalesce (at $t$=80). The time to coalescence
approaches infinity as the $\mathcal{PT}$ transition is approached.
The phase delay therefore can only be properly defined in the unstable regime,
where for $k\rightarrow 0$ it is equal to $\pi/2$ right at the $\mathcal{PT}$ transition
and zero far from the transition point $\tau_{crit}$, bearing in mind that for small
$k$ the parameter space for stability is tiny (Fig.~\ref{kgammaG}a).
As with the case of broken $\mathcal{PT}-$symmetry in the overdamped limit of the 
classical dimer, for $V<\gamma$ all of the solutions diverge monotonically for a constant 
gain/loss $\gamma$.
\begin{figure}[h]
\includegraphics[width=8cm]{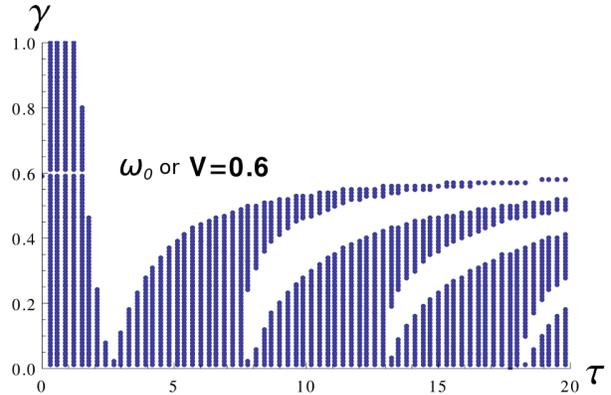}
\caption{Stability phase diagram on the $\gamma-\tau$ parameter plane for 
the single oscillator, the classical dimer with $k=0$ or the linear Schr\"{o}dinger
dimer for ($V$ or $\omega_0$)=0.6.
The shaded (blank) regions are stable (unstable).}
\label{schrophase}
\end{figure}

\subsection{Time-dependent gain/loss function}
For the time-alternating $\gamma(t)$ of Eq.~\ref{gainloss}, $\Psi_a$ and $\Psi_b$ 
can be obtained analytically using the same methods described earlier,
and a map that connects the $\Psi$s at times $t$ and $t+2\tau=t+T$ 
can be constructed so that
\begin{equation}
\begin{pmatrix}
  \Psi_a\\
  \Psi_b
\end{pmatrix}=\underline{\underline{M}}_{G/L}(\tau)
\begin{pmatrix}
  {\Psi_{a,0}}  \\
  {\Psi_{b,0}}
\end{pmatrix}
\label{Mmatrix2}
\end{equation}
where $\Psi_{a,0} =\Psi_{a} (t=0)$ and $\Psi_{b,0} =\Psi_{b} (t=0)$.
Examining the stability of the propagation matrix in Eq.~\ref{Mmatrix2}, we find 
a direct correspondence with the behaviour of the single classical oscillator, 
mainly a regime $V<\gamma$ with diverging orbits beyond a certain $\tau$ where the 
instability criterion is established by the divergence of the eigenvalue magnitudes at
\begin{equation}
  \tau_{crit}=\frac{\arccos{(\gamma/V)}}{\delta} ,
  \label{instability2}
\end{equation}
where $\delta=\sqrt{V^2-\gamma^2}$, and a regime with $V>\gamma$ containing
periodic $\mathcal{PT}$-symmetry breaking/recoveries at
\begin{equation}
   \cos{(2\delta \tau)}=\left( \frac{2\gamma^2}{V^2}-1\right) .
   \label{instability3}
\end{equation}
\begin{figure}[h]
\includegraphics[width=8cm]{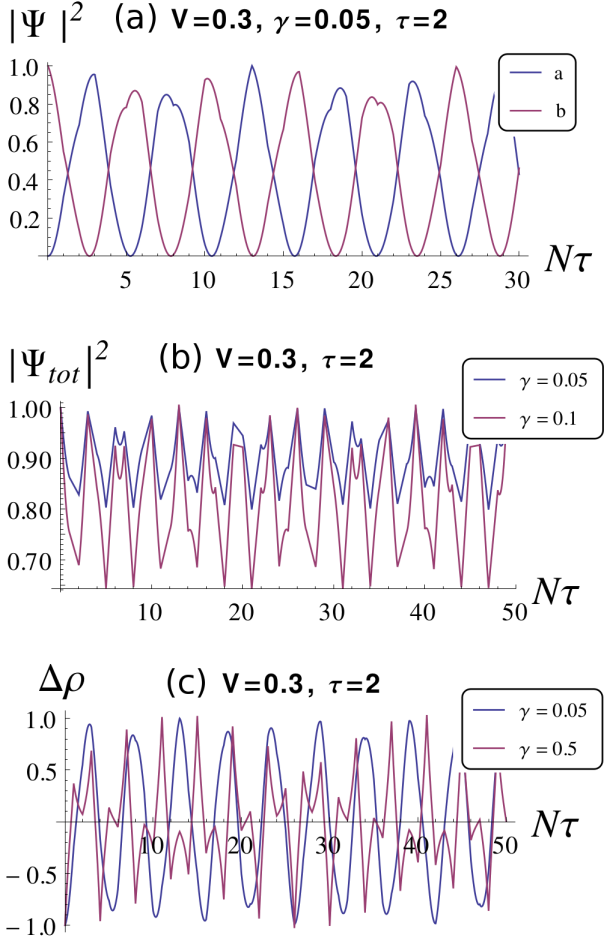}
\caption{Time-dependence of 
(a) the probabilities $\left|\Psi_a(t)\right|^2$ and $\left|\Psi_b(t)\right|^2$
    for $\gamma<V$; 
(b) the total probability $P$ for $\gamma<V$;
(c) the probability difference $\Delta\rho$ for $\gamma<V$ and $\gamma>V$. 
The initial conditions in all cases are $\Psi_a(0)=0$, $\Psi_b(0)=1$.
In all cases shown the system is in the exact $\mathcal{PT}-$symmetric phase.}
\label{dimertraj}
\end{figure}

The overall $\gamma-\tau$ phase diagram is shown in Fig.~\ref{schrophase} for $V$=0.6. 
For values of $\gamma<V$ there is an infinite series of $\mathcal{PT}-$symmetry
breaking/recoveries while for $\gamma>V$ there is just a single, small region of 
stability. The probabilities for the oscillators $a$ and $b$ of the LSD in
the exact $\mathcal{PT}$-symmetric phase are shown for a case of $\gamma<V$ in 
Fig.~\ref{dimertraj}a, while their difference 
\begin{equation}
\label{difference}
  \Delta\rho =\left|\Psi_a (t)\right|^2 -\left|\Psi_b (t)\right|^2 
\end{equation}
is shown in Fig.~\ref{dimertraj}c for both $\gamma<V$ and $\gamma>V$.
While the probability of each component certainly is expected to oscillate in time,
the non-conservation of the total probability is solely a feature of the non-Hermiticity
of the system and it is more pronounced for a larger gain/loss parameter $\gamma$
(Fig.~\ref{dimertraj}b). It occurs because when Hermiticity is violated,
$\Psi_a$ and $\Psi_b$ no longer form an orthonormal basis of the non-Hermitian 
Hamiltonian in Eq. (\ref{hschro}) \cite{Zheng2010}. The oscillation of $\Delta\rho$,
on the other hand, is bounded in the exact $\mathcal{PT}$-symmetric phase 
(Fig.~\ref{dimertraj}c) by $\pm 1$, irrespective of the value of $\gamma$.

\section{Conclusions}
We have investigated theoretically two $\mathcal{PT}$-symmetric dimer systems
which are representative of two wide classes of models used for physical 
systems, and we have made comparisons to recent experiments covering both of 
these classes. In the first class considered, based on classical  oscillators,
all of the parameters are real and it is typically concerned with observables
such as \textit{e.g.} position, charge, etc., while the second, describing 
the time-evolution of field amplitudes, pertains to systems with generally 
complex parameters (\textit{e.g.} index of refraction) and it is typically
by optics experiments.  ${\cal PT}-$symmetric dimer systems fabricated in
a laboratory may be adequately modeled by one of the two models presented above.
Thus, it is imperative to be able to predict their behavior in particular 
regions of parameter space. Phase diagrams, in which stability/instability
regions corresponding to the exact/broken ${\cal PT}-$symmetric phase of
the systems are shown, provide valuable information in that respect.
In particular, they may be very helpful in identifying regions of the parameter
space where stable solutions exist.

In a similar manner to how re-entrant ${\cal PT}-$symmetry transitions have been 
predicted in single classical oscillator with a periodically varied gain/loss
parameter~\cite{Tsironis2014}, the coupling parameter $k$ of the classical dimer
induces another layer of ${\cal PT}-$symmetry transitions. For small values
of $k$ the phase space for re-entrant ${\cal PT}-$symmetry transitions is reduced,
while above a critical value $k_{crit}$ the transitions return and the system 
achieves greater stability as the coupling is increased beyond this point.
Note that ${\cal PT}-$symmetry transitions as a function of the coupling strength 
have been recently observed in a coupled microresonator system \cite{Peng2014}.
The solutions of the two dimer models considered here display similar characteristics,
such as a phase shift between the oscillator or field amplitudes correspondingly,
as well as a non-conservation of oscillator or probability amplitude in the exact
${\cal PT}-$symmetric phase, and this distinguishes them from conventional (undamped)
stability.

It is important to note that the two dimer models only reduce to the behaviour 
of a single classical oscillator, or "zero-dimensional" ${\cal PT}-$symmetric 
system \cite{Tsironis2014} that we chose as a benchmark here,
under very specific and different conditions: paths, \textit{i.e.} slices of phase
space, and order of limits taken. As well, the two dimer systems may have very 
similar behaviour but they display subtle differences, particularly in the case of a 
constant gain/loss function $\gamma$ studied in great detail here.
In order to make their behaviour coincide, limiting approaches need to be taken
in a specific, and opposite sequence. The path-dependent and/or limiting behaviour 
is very important to be aware of in making correspondences between classical or
wave-based models or in comparing or interpreting results from experiments belonging
to the two different classes.
As these simple dimer systems can be used as basic elements for the fabrication of
larger networks comprised of dimers as in Refs.
\cite{Regensburger2012,Lazarides2013a,Konotop2012,Suchkov2011}, understanding the 
properties of the dimer elements is invaluable for deducing the properties of the
system as a whole.

\begin{acknowledgement}
This work was partially supported by
the European Union's Seventh Framework Programme (FP7-REGPOT-2012-2013-1)
under grant agreement n$^o$ 316165, and
by the Thales Project ANEMOS, co‐financed by the European Union
(European Social Fund – ESF) and Greek national funds through the Operational
Program "Education and Lifelong Learning" of the National Strategic Reference
Framework (NSRF) ‐ Research Funding Program: THALES.
Investing in knowledge society through the European Social Fund.
\end{acknowledgement}

\bibliographystyle{IEEEtran}

\begin{thebibliography}{99}

\bibitem{Hook2012}
  Hook, D.W.: 
  Non-Hermitian potentials and real eigenvalues.
\newblock Ann. Phys. (Berlin) \textbf{524 (6-7)}, A106 (2012)
(and references therein)

\bibitem{ElGanainy2007}
  El-Ganainy, R., Makris, K.G., Christodoulides, D.N., Musslimani, Z.H.: 
  Theory of coupled optical $\cal PT-$symmetric structures.
\newblock Opt. Lett. \textbf{32}, 2632--2634 (2007)

\bibitem{Makris2008}
  Makris, K.G., El-Ganainy, R., Christodoulides, D.N., Musslimani, Z.H.: 
  Beam dynamics in $\cal PT-$symmetric optical lattices.
\newblock Phys. Rev. Lett. \textbf{100}, 103904 (2008)

\bibitem{Guo2009}
  Guo, A., Salamo, G.J., Duchesne, D., Morandotti, R., Volatier-Ravat, M.,
  Aimez, V., Siviloglou, Christodoulides, D.N.:
  Observation of $\cal PT-$symmetry breaking in complex optical potentials.
  \newblock Phys. Rev. Lett. \textbf{103}, 093902 (2009)

\bibitem{Ruter2010}
  R{\"u}ter, C.E., Makris, K.G., El-Ganainy, R., Christodoulides, D.N., Segev,
  M., Kip, D.: 
  Observation of parity-time symmetry in optics.
\newblock Nature Physics \textbf{6}, 192-- (2010)

\bibitem{Regensburger2012}
  Regensburger, A., Bersch, C., Miri., M.-A., Onishchukov, G., Christodoulides, D. N.,
  Peschel, U.:
  Parity-time synthetic photonic lattices.
\newblock Nature  \textbf{488}, 167 (2012)

\bibitem{Feng2012}
  Feng, L., Xu, Y.L., Fegadolli, W.S., Lu, M.-H., Oliveira, J.E.B., Almeida, V.R.,
  Chen, Y.-F., Scherer, A.:
  Experimental demonstration of a unidirectional reflectionless parity-time
  metamaterial at optical frequnecies.
\newblock Nature Materials \textbf{12}, 108-113 (2013)

\bibitem{Feng2014}
  Feng, L., Wong, Z.J., Ma, R., Wang, Y., Zhang, X.:
  Parity-time synthetic laser.
\newblock arXiv:1405.2863

\bibitem{Peng2014}
  Peng, B., {\"O}zdemir, S.K., Lei, F., Monifi, F., Gianfreda, M., Long, G. L.,
  Fan, S., Nori, F., Bender, C.M., Yang, L.: 	
  Parity–time-symmetric whispering-gallery microcavities.
  Nature Physics \textbf{10}, 394-398 (2014)




\bibitem{Dmitriev2010}
  Dmitriev, S.V., Sukhorukov, A.A., Kivshar, Y.S.: 
  Binary parity-time-symmetric nonlinear lattices with balanced gain and loss.
\newblock Opt. Lett. \textbf{35}, 2976--2978 (2010)

\bibitem{Li2011}
  Li, K., Kevrekidis, P.G.: $\cal PT-$symmetric oligomers: 
  Analytical solutions, linear stability, and nonlinear dynamics.
\newblock Phys. Rev. E \textbf{83}, 066608 (2011)

\bibitem{Ambroise2012}
  D'Amroise, J.D., Kevrekidis, P.G., Lepri, S.:
  Asymmetric wave propagation through nonlinear $\cal PT-$symmetric oligomers.
\newblock  J. Phys. A: Math. Theor. \textbf{45}, 444012 (2012)


\bibitem{Schindler2011}
  Schindler, J., Li, A., Zheng, M.C., Ellis, F.M., Kottos, T.: 
  Experimental study of active LRC circuits with $\cal PT$ symmetries.
\newblock Phys. Rev. A \textbf{84}, 040101(R) (2011)

\bibitem{Bender2013}
  Bender, N., Factor, S., Bodyfelt, J.D., Ramezani, H., Christodoulides, D.N.,
  Ellis, F.M., Kottos, T,:
  Observation of asymmetric transport in structures with active nonlinearities.
\newblock Phys. Rev. Lett.  \textbf{110}, 234101 (2013)

\bibitem{Lazarides2013a}
  Lazarides, N., Tsironis, G.P.:
  Gain-driven discrete breathers in ${\cal PT-}$symmetric nonlinear metamaterials.
\newblock Phys. Rev. Lett.  \textbf{110}, 053901 (2013)

\bibitem{Tsironis2014}
  Tsironis, G.P., Lazarides, N.:
  ${\cal PT-}$symmetric nonlinear metamaterials and zero-dimensional systems.
\newblock Appl. Phys. A \textbf{115}, 449--458 (2014)

\bibitem{Sun2014}
  Sun, Y., Tan, W., Li, H.-Q., Li, J., Chen, H.:
  Experimental demonstration of a coherent perfect absorber with ${\cal PT}$
  phase transition.
\newblock Phys. Rev. Lett. \textbf{112}, 143903 (2014)

\bibitem{Ramezani2010}
  Ramezani, H., Kottos, T., El-Ganainy, R., Christodoulides, D.N.:
  Unidirectional nonlinear ${\cal PT-}$symmetric optical structures.
  \newblock  Phys. Rev. A \textbf{82}, 043803 (2010) 

\bibitem{Benisty2011}
  Benisty, H., Degiron, A., Lupu, A., De Lustrac, A., Ch{\'e}nais, S.,
  Forget,i S., Besbes, M., Barbillon, G., Bruyant, A., Blaize, S.,
  L{\'e}rondel, G.:
  Implementation of $\cal PT-$symmetric devices using plasmonics: principle
  and applications.
\newblock Opt. Express \textbf{19}, 18004 (2011)

\bibitem{Barashenkov2012}
  Barashenkov, I.V., Suchkov, S.V., Sukhorukov, A.A., Dmitriev, S.V., Kivshar, Yu.S.:
  Breathers in $\cal PT-$symmetric optical couplers.
\newblock  Phys. Rev. A \textbf{86}, 053809 (2012)

\bibitem{Lupu2013}
  Lupu, A., Benisty, H., Degiron, A.:
  Switching using $\cal PT-$symmetry in plasmonic systems: positive role of the losses.
\newblock Opt. Express \textbf{21}, 21651 (2013)

\bibitem{Duanmu2013}
  Duanmu, M., Li, K., Horne, R.L., Kevrekidis, P.G., Whitaker, N.:
  Linear and nonlinear parity-time-symmetric oligomers: a dynamical systems analysis.
\newblock Phil Trans R Soc A \textbf{371}, 20120171 (2013) 

\bibitem{Cuevas2013}
  Cuevas, J., Kevrekidis, P.G., Saxena, A., Khare, A.:
  $\cal PT-$symmetric dimer of coupled nonlinear oscillators.
\newblock Phys. Rev. A \textbf{88}, 032108 (2013)

\bibitem{Barashenkov2013}
  Barashenkov, I.V., Jackson, G.S., Flach, S.:
  Blow-up regimes in the $\cal PT-$symmetric coupler and the actively coupled dimer.
\newblock Phys. Rev. A \textbf{88}, 053817 (2013)

\bibitem{Molina2014}
  Molina, M.I.:
  Bounded dynamics in finite $\cal PT-$symmetric magnetic metamaterials.
\newblock Phys. Rev. A \textbf{89}, 033201 (2014)

\bibitem{Alexeeva2014}
  Alexeeva, N.V., Barashenkov, I.V., Rayanov, K., Flach, S.:
  Actively coupled optical waveguides.
\newblock  Phys. Rev. A \textbf{89}, 013848 (2014)

\bibitem{Bludov2012}
  Bludov, Yu.V., Konotop, V.V., Malomed, B.A.:
  Stable dark solitons in $\cal PT-$symmetric dual-core waveguides.
\newblock Phys. Rev. A \textbf{87}, 013816 (2013)

\bibitem{Bludov2013}
  Bludov, Yu.V., Driben, R., Konotop, V.V., Malomed, B.A.:
  Instabilities, solitons an rogue waves in $\cal PT-$coupled nonlinear 
  waveguides. 
\newblock J. Opt. \textbf{15}, 064010 (2013)

\bibitem{Konotop2012}
  Konotop, V.V., Pelinovsky, D.E., Zezyulin, D.A.:
  Discrete solitons in $\cal PT-$symmetric lattices.
\newblock EPL \textbf{100}, 56006 (2012)  

\bibitem{Pickton2013}
  Pickton, J., Susanto, H.:
  Integrability of ${\cal PT}$-symmetric dimers.
\newblock Phys. Rev. A  \textbf{88}, 063840 (2013)

\bibitem{Zheng2010}
  Zheng, M.C., Christodoulides, D.N., Fleischmann, R., Kottos, T.:
  ${\cal PT}$ optical lattices and universality in beam dynamics.
\newblock Phys. Rev. A  \textbf{82}, 010103(R) (2010)

\bibitem{Suchkov2011}
  Suchkov, S.V., Malomed, B.A., Dmitriev, S.V., Kivshar, Y.S.: 
  Solitons in a chain of parity-time invariant dimers.
\newblock Phys. Rev. A \textbf{84}, 046609 (2011)

\end{thebibliography}

\end{document}